\documentclass{osa-article}

\journal{oe}

\articletype{Research Article}
\usepackage{soul}
\usepackage[numbers,sort&compress]{natbib}
\begin{document}

\title{Confocal super-resolution microscopy based on a spatial mode sorter}

\author{Katherine K. M. Bearne,\authormark{1,$\dagger$} Yiyu Zhou,\authormark{2,$\dagger$,*} Boris Braverman,\authormark{1} Jing Yang,\authormark{3} S. A. Wadood,\authormark{2} Andrew N. Jordan,\authormark{3,4} A. N. Vamivakas,\authormark{2,3,5} Zhimin Shi,\authormark{6} and Robert W. Boyd\authormark{1,2,3} }
\address{\authormark{1}Department of Physics, University of Ottawa, Ottawa, Ontario K1N 6N5, Canada\\
\authormark{2}The Institute of Optics, University of Rochester, Rochester, New York 14627, USA\\
\authormark{3}Department of Physics and Astronomy, University of Rochester, Rochester, New York 14627, USA\\
\authormark{4}{Institute for Quantum Studies, Chapman University, Orange, California 92866, USA}\\
\authormark{5}{Materials Science Program, University of Rochester, Rochester, New York 14627, USA}\\
\authormark{6}Department of Physics, University of South Florida, Tampa, Florida 33620, USA\\
\authormark{$\dagger$}These authors contributed equally}

\email{\authormark{*}yzhou62@ur.rochester.edu}



\begin{abstract}
Spatial resolution is one of the most important specifications of an imaging system. Recent results in quantum parameter estimation theory reveal that an arbitrarily small distance between two incoherent point sources can always be efficiently determined through the use of a spatial mode sorter. However, extending this procedure to a general object consisting of many incoherent point sources remains challenging, due to the intrinsic complexity of multi-parameter estimation problems. Here, we generalize the Richardson-Lucy (RL) deconvolution algorithm to address this challenge. We simulate its application to an incoherent confocal microscope, with a Zernike spatial mode sorter replacing the pinhole used in a conventional confocal microscope. We test different spatially incoherent objects of arbitrary geometry, and we find that the resolution enhancement of sorter-based microscopy is on average over 30\% higher than that of a conventional confocal microscope using the standard RL deconvolution algorithm. Our method could potentially be used in diverse applications such as fluorescence microscopy and astronomical imaging.
\end{abstract}

\section{Introduction}

Enhancing spatial resolution is a persistent goal for imaging systems. The resolution of an incoherent far-field imaging system was previously believed to be limited by Rayleigh's criterion \cite{rayleigh1879xxxi}{}. In recent decades, a multitude of super-resolution methods have been demonstrated to break the diffraction limit, such as stimulated-emission depletion (STED) \cite{hell1994breaking}{}, photoactivated localization microscopy (PALM) \cite{betzig2006imaging}{}, and stochastic optical reconstruction microscopy (STORM) \cite{rust2006sub}{}. However, these methods generally require the use of specially prepared fluorescent molecules, and the data collection in an experiment can take a long time. In addition to these classical methods, various quantum effects have been investigated to enhance the imaging resolution. Optical centroid measurement \cite{tsang2009quantum, shin2011quantum, rozema2014scalable, unternahrer2018super, toninelli2019resolution}{} is another quantum approach that can improve the resolution by up to 41\% via detecting the centroid of entangled bi-photons. The anti-bunching effect has also been exploited to enhance the spatial resolution when imaging single-photon sources such as quantum dots through the use of coincidence measurement \cite{tenne2019super, schwartz2013superresolution}{}. Nonetheless, these non-classical methods typically require the use of quantum, low-brightness light sources (e.g., single-photon sources and entangled-photon sources) as well as slow, high-order intensity correlation measurements, which limits their widespread adoption in real-world imaging systems.


\begin{figure}[t]
\centering
\includegraphics[width=\linewidth]{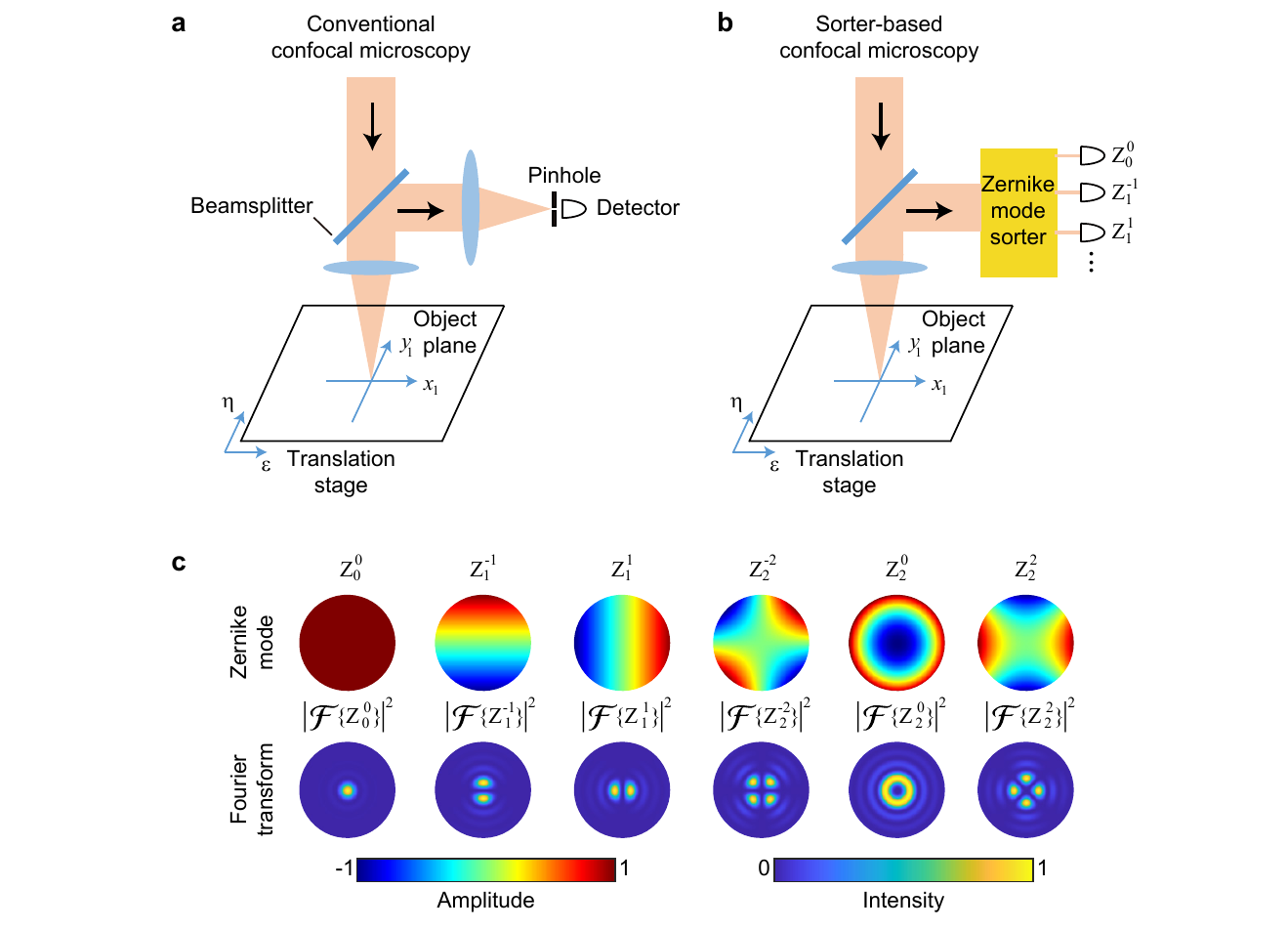}
\caption{(a) Schematic of a conventional confocal microscope. (b) Schematic of the sorter-based confocal microscopy. Conventional confocal microscopy uses a pinhole and a single detector, while the proposed scheme uses a spatial mode sorter to first decompose the received field, with every output port of the sorter measured by a separate detector. (c) The first six Zernike modes $Z_n^m$ and the intensity profiles of their respective Fourier transforms $|\mathcal{F}\{Z_n^m\}|^2$.}
\label{fig:figure1}
\end{figure}

In recent years, a quantum-inspired super-resolution imaging method based on spatial mode sorting (called SPADE) has been proposed \cite{tsang2016quantum}{} and experimentally demonstrated \cite{yang2016far, paur2016achieving, tham2017beating, tang2016fault, zhou2019quantum}{}. The Rayleigh diffraction limit can be broken through the use of an appropriate spatial mode sorter, and an arbitrarily small separation between two spatially incoherent, equally bright point sources can be well resolved. Although the theory for SPADE was developed in the framework of quantum metrology, it can be interpreted classically \cite{tsang2018subdiffraction}{} and does not need non-classical light sources or high-order coincidence detection, which is the major advantage over the aforementioned super-resolution methods. The theoretical treatment of SPADE approaches the super-resolution task as a parameter estimation problem, which works well when imaging a scene with a small number of point objects. However, it is non-trivial to apply the theory to a general scene that contains many point sources or continuous objects due to the complexity of multi-parameter estimation problems \cite{albarelli2019evaluating, tsang2019semiparametric}{}. A few theoretical attempts have been made towards sorter-based super-resolution imaging for a scene with a very small number of unknown parameters \cite{zhou2019modern, yang2019optimal, tsang2019semiparametric, tsang2017subdiffraction, vrehavcek2017multiparameter, lu2018quantum, prasad2020quantum, bonsma2019realistic, tsang2020efficient, peng2020generalization}{}. However, these previous works mainly focus on the Fisher information analysis of the mode sorter, which exhibits intractable complexity in the calculation of quantum Fisher information. Furthermore, even if the quantum Fisher information can be computed, it is challenge to determine if the quantum Fisher information can be achieved by practical measurements for all parameters. Hence, to the best of our knowledge, no method has yet been reported to super-resolve an object of arbitrary geometry using a mode sorter. Here we address this challenge by treating the sorter-based super-resolution imaging as a deconvolution problem. We propose to replace the pinhole in a standard confocal microscope with a spatial mode sorter. We generalize the standard RL deconvolution algorithm \cite{richardson1972bayesian, lucy1974iterative}{} to digitally process the multiple outputs of the mode sorter in order to reconstruct a super-resolved image. In Section~\ref{sec:algorithm}, we introduce the conceptual schematic of the sorter-based confocal microscopy as well as the algorithm for image reconstruction. In Section~\ref{sec:simulation} we present the numerical simulation results. The conclusion of this work is discussed in Section~\ref{sec:conclusion}.

\section{Generalized Richardson-Lucy deconvolution algorithm}\label{sec:algorithm}

The schematic of a confocal microscope is shown in Fig.~\ref{fig:figure1}(a,b). Conventional confocal microscopy uses a pinhole in the image plane before the single-pixel detector. Here we assume that the illumination beam is spatially coherent and the light scattered by the object is spatially incoherent, which is common in fluorescence microscopy. The illumination and reflected beam wavelengths are assumed to be the same for simplicity, although they can be different in fluorescence microscopy. We use $W_0$ to describe the object brightness profile and use $M$ to denote the point spread function (PSF) of the conventional confocal microscope using a pinhole of diameter $D$. Here we assume a circular aperture of the objective lens, and the PSF of the objective lens is thus an Airy disk \cite{airy1835diffraction}{}. Additional details of $M$ are presented in Supplementary Section~1. By raster scanning the object, a 2D image can be obtained, and the resultant confocal image $I_{\rm{con}}$ can be described by the convolution of $W_0$ and $M$ as $I_{\rm{con}}=M*W_0$. In our model we consider only the fundamental quantum noise which leads to Poissonian photon statistics; we ignore other technical sources of noise. Therefore, the shot-noise-limited image that can be experimentally measured is described by $I_{\rm{con}}^{\rm{exp}}={\rm{Poisson}}(I_{\rm{con}})$, where ${\rm{Poisson}(\cdot)}$ denotes one random realization of the Poisson distribution for a given mean. The standard RL deconvolution algorithm can be expressed as \cite{richardson1972bayesian, lucy1974iterative}{}
\begin{equation}
\begin{aligned}
W_{r+1} =W_{r} \cdot \left( M * \frac{I_{\rm{con}}^{\rm{exp}}}{M*W_r} \right) ,
\end{aligned}
\label{eq:convRL}
\end{equation}
where $W_r$ is the deconvolved image in the $r$-th iteration and $*$ denotes convolution. In general, the iterative deconvolution algorithm begins with an image of uniform intensity $W_{r=1}={\rm{const}}$, and the term inside the parentheses in the above equation can be understood as a correction to $W_r$ during each iteration. The proposed sorter-based confocal microscope is shown in Fig.~\ref{fig:figure1}(b). It can be seen that a Zernike mode sorter is used in the Fourier plane. Here Zernike modes are adopted because they have been shown to be the optimal basis for an imaging system with a circular aperture \cite{yang2019optimal, yu2018quantum}{}. In particular, we choose the six lowest-order Zernike modes ($Z_0^0$, $Z_{1}^{-1}$ , $Z_{1}^{1}$ , $Z_{2}^{-2}$ , $Z_{2}^0$, and $Z_2^2$), as shown in Fig.~\ref{fig:figure1}(c). The Zernike mode sorter projects the collected photons onto each Zernike mode $Z_n^m$ in the Fourier plane, and each output port of the Zernike mode sorter produces a 2D image $H_{mn}$ by raster scanning the object. The image $H_{mn}$ is given by the convolution of the original object image $W_0$ and $Q_{mn}$ as $H_{mn}=W_0 * Q_{mn}$, where $Q_{mn}$ is the effective PSF when projecting into the mode $Z_n^m$:
\begin{equation}
\begin{aligned}
Q_{mn}(x_1,y_1) &= N_0 \frac{k^2 \text{NA}^2}{4\pi} B_{m=0,n=0}(r_1,\theta_1) \cdot B_{mn}(r_1,\theta_1), \\
 B_{mn}(r_1,\theta_1)&=\frac{8(n+1)}{\epsilon_m } \frac{J^2_{n+1}(k \text{NA} r_1)}{(k\text{NA}r_1)^2}\sin^2(m\theta_1 + \frac{\pi}{2}\cdot {\mathcal{H}}(m)),
\end{aligned}
\end{equation}
where $J_{n+1}(\cdot)$ is the Bessel function of order $n+1$; $\epsilon_m = 2$ if $m=0$ and $\epsilon_m = 1$ if $m\neq 0$; ${\mathcal{H}}(m)$ is the Heaviside step function where ${\mathcal{H}}(m)=1$ if $m \geqslant 0$ and ${\mathcal{H}}(m)=0$ if $m < 0$; $(x_1,y_1)$ are the Cartesian coordinates at the object plane, and $(r_1,\theta_1)$ are the corresponding polar coordinates; $N_0$ is the photon number in the illumination beam at each raster scanning step; $k=2\pi/\lambda$ is the wave number, $\lambda$ is the wavelength, and NA is the collection numerical aperture of the objective lens. In this equation, $B_{m=0,n=0}$ represents the PSF of the illumination beam, and $B_{mn}$ is the intensity profile of the Fourier transform of $Z_n^m$ in the image plane as shown in the bottom row in Fig.~\ref{fig:figure1}(c). Here, we assume that the illumination beam has a flat spatial profile before being focused by the objective lens and that the illumination NA is the same as the collection NA. We use this assumption to simplify the calculation and simulation, but we note that this assumption can be relaxed and is not necessary to the result. Derivations of the analytical form of $B_{mn}$, $H_{mn}$ and $Q_{mn}$ are presented in Supplementary Section~2. We use $H_{mn}^{\rm{exp}}={\rm{Poisson}}(H_{mn})$ to denote the randomly generated shot-noise-limited image that can be measured in an experiment. As one can see, a major difference between the conventional confocal microscopy and the sorter-based confocal microscopy is that multiple images (six in our case) can be obtained simultaneously when using a mode sorter. Here we propose a generalized RL deconvolution algorithm, which can be expressed as
\begin{equation}
\begin{aligned}
W_{r+1} =W_{r} \cdot \sum_{mn} \left( Q_{mn} * \frac{H_{mn}^{\text{exp}}}{Q_{mn}*W_r} \right).
\label{eq:genRL}
\end{aligned}
\end{equation}
Compared to the conventional RL deconvolution algorithm (Eq.~(\ref{eq:convRL})), it can be seen that the correction term inside the parentheses in the above equation is the sum of contributions from different modes.

\section{Numerical simulation}\label{sec:simulation}
\subsection{Algorithm performance evaluation}

\begin{figure}[t]
\centering
\includegraphics[width=\linewidth]{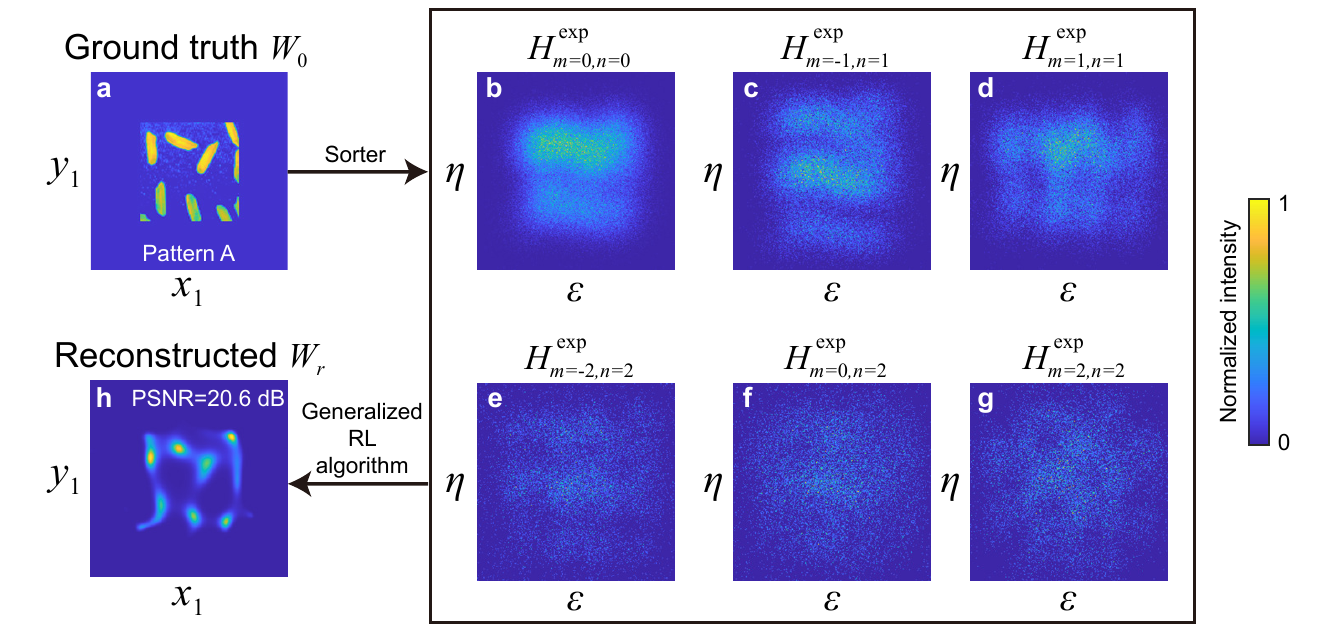}
\caption{Example of image reconstruction with the sorter-based super-resolution approach. (a) The ground truth image. (b-g) The 2D images $H_{mn}^{\rm{exp}}$ that can be measured via a Zernike mode sorter by raster scanning the ground truth image at the object plane in the presence of Poisson noise. (h) The reconstructed super-resolved image obtained by feeding the data $H_{mn}^{\rm{exp}}$ into the generalized RL deconvolution algorithm.}
\label{fig:figure2}
\end{figure}

\begin{figure*}[t]
\centering
\includegraphics[width=\linewidth]{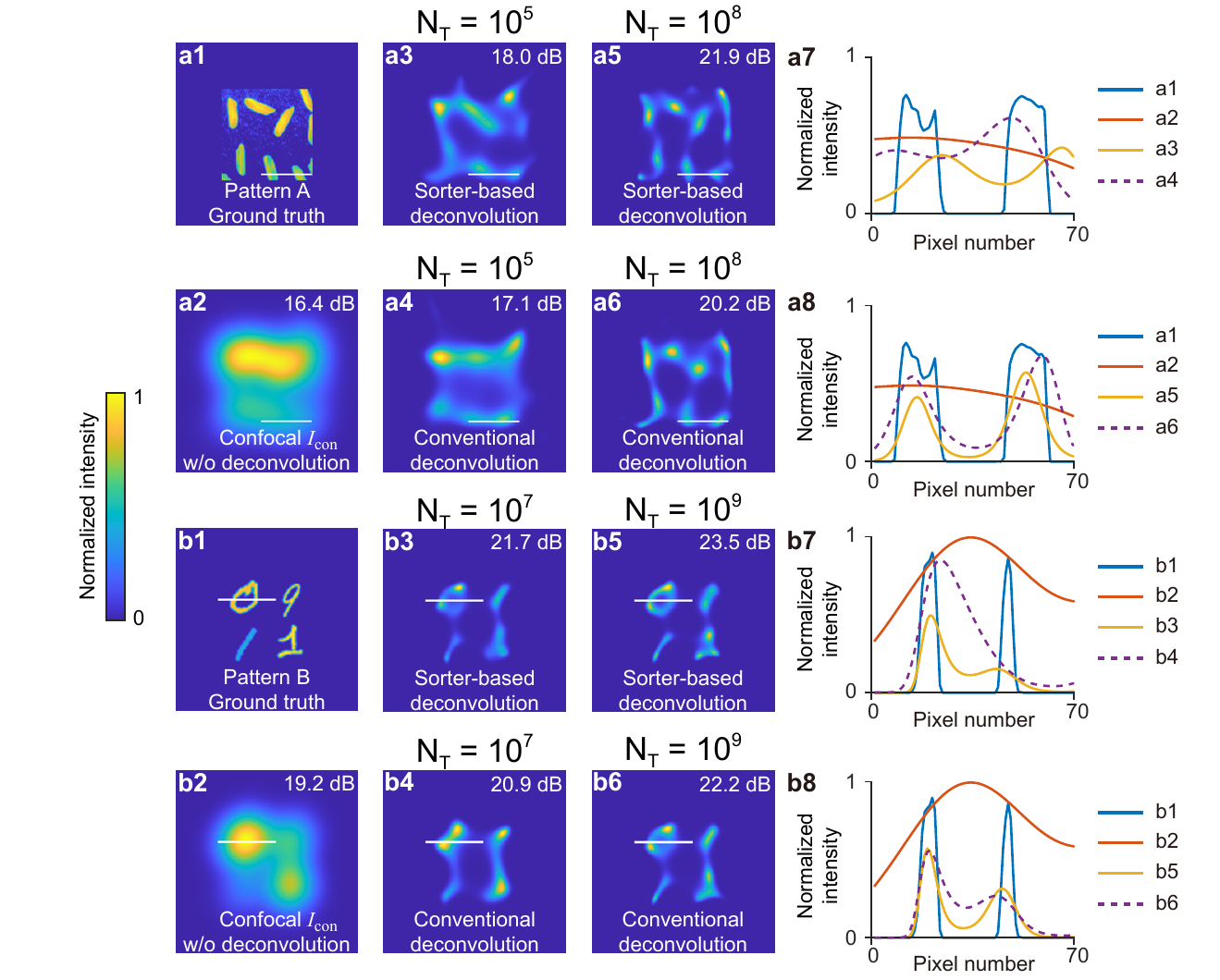}
\caption{(a1) Ground truth image for pattern A. (a2) The confocal diffraction-limited image without deconvolution. (a3-a6) The reconstructed images. The used algorithm and the total photon number $N_T$ in the illumination beam are labeled near each image. (a7, a8) The 1D cross-section lines for (a1-a6) indicated by the corresponding white bars. (b1-b8) Results for pattern B. The PSNR is shown at the upper right corner of each image. In all 1D cross-section lines, it can be seen that the sorter-based deconvolution algorithm consistently shows better contrast and fidelity to the ground truth than the conventional deconvolution algorithm.}
\label{fig:figure3}
\end{figure*}

We next present the results of numerical simulations that implement the generalized deconvolution algorithm and compare its performance to that of the conventional deconvolution algorithm. One of the objects we use (pattern A) is shown in Fig.~\ref{fig:figure2}(a). The original object image has a size of 128$\times$128 pixels and is zero padded to 256$\times$256 pixels to avoid the diffraction-induced boundary clipping effect. A 2D image $H^{\exp}_{mn}$ can be obtained at each output port of the mode sorter when raster scanning the translation stage by $(\epsilon, \eta)$. Here we choose the scanning step size to be 1 pixel and the total scanning steps to be $256\times 256$, resulting in a 2D image $H^{\exp}_{mn}(\epsilon, \eta)$ of $256\times 256$ pixels. At each scanning step, we assume that $N_0$ photons are used to illuminate the object, and thus the total photon number in the illumination beam is $N_T=256\times 256 \times N_0$. In this work, we use $N_T$ as a variable and perform simulations under different $N_T$. This is because $N_T$ is typically controllable in an experiment by adjusting the illumination laser power and is independent of the sample properties. The results for different Zernike mode outputs are shown in Fig.~\ref{fig:figure2}(b-g). One can see that the output of high-order modes has a lower photon count and is thus more susceptible to Poisson noise. We use $W_{r=1}={\rm{const}}$ as the starting point and run the iterative deconvolution algorithm based on Eq.~(\ref{eq:genRL}). We choose the commonly used peak signal-to-noise ratio (PSNR) to quantify the quality of the reconstructed image $W_r$. The definition of PSNR is given by \cite{nasrollahi2014super}{}
\begin{equation}
\begin{aligned}
{\rm{PSNR}} =10 \log_{10} \frac{\max(W_0)^2}{\frac{1}{N^2} \sum_{i=1}^{N}\sum_{j=1}^{N} \left|W_0(i,j)-W_r(i,j)\right|^2},
\end{aligned}
\label{eq:PSNR}
\end{equation}
where $(i,j)$ are the integer pixel indices of the digital image and $N=256$ is the pixel size along one dimension. We stop the deconvolution algorithm at a maximum iteration number $N_{\rm{ite}}= 10^4$, which is limited by time and computational power constraints. In general, the PSNR increases with increasing iteration number $r$. However, if the data is noisy, the noise can be amplified when the iteration number is large, and thus the PSNR can decrease if $r$ continues to increase. In our implementation, we monitor the PSNR as a function of the iteration $r$ and choose the maximum PSNR for $1 \leqslant r \leqslant N_{\rm{ite}}$ for each implementation. The reconstructed image is shown in Fig.~\ref{fig:figure2}(h). More details on the PSNR as a function of the iteration number are provided Supplementary Section~3. For practical applications where the ground truth is not available, a stopping criterion \cite{herbert1990statistical} must be used. The simplest (and perhaps the most widely used) stopping criterion is to manually specify a maximum iteration number. The relation between PSNR and the iteration number for both sorter-based deconvolution algorithm and the conventional deconvolution algorithm is presented in Supplementary Section~3 to illustrate the effect of a manually specified stopping criterion. The results show that both the conventional and sorter-based deconvolution algorithms have a similar dependence on the iteration number, and thus the PSNR improvement is almost independent of the chosen stopping criterion.

We next characterize the performance of the generalized deconvolution algorithm under different levels of Poisson noise by adjusting the total photon number $N_T$ in the illumination beam. The ground truth for pattern~A is shown in Fig.~\ref{fig:figure3}(a1), and we choose the Rayleigh-criterion resolution $\delta x_0=1.22 \pi /(k \rm{NA})=0.61 \lambda / \text{NA}$ to be 80 pixels. We emphasize that only the relative ratio between $\lambda / \text{NA}$ and the pixel pitch size is important, and here we do not specify the respective value of these parameters for generality. The noiseless, diffraction-limited confocal image without deconvolution $I_{\rm{con}}$ is shown in Fig.~\ref{fig:figure3}(a2), which is too blurry to reveal the details of the ground truth. We next vary the total photon number $N_T$ and test the performance of the deconvolution algorithm with different $N_T$. The reconstructed images by the sorter-based deconvolution algorithm and the conventional deconvolution algorithm are presented in Fig.~\ref{fig:figure3}(a3-a6). We also test another pattern B made of four handwritten digits (MNIST handwritten digit database \cite{lecun1998gradient}{}) with non-uniform intensity profile as shown in Fig.~\ref{fig:figure3}(b1). The conventional confocal image $I_{\rm{con}}$ without deconvolution is shown in Fig.~\ref{fig:figure3}(b2), and the digits cannot be resolved based on this image. The sorter-based deconvolved images are presented in Fig.~\ref{fig:figure3}(b3,b5), and the conventional deconvolved images are presented in Fig.~\ref{fig:figure3}(b4,b6). It can be seen that the digits `0' and `1' using the sorter-based approach are more visually resolvable that the conventional deconvolved results. Fig.~\ref{fig:figure3}(a7,a8,b7,b8) are 1D cross-sections through the images (indicated by the white bars), comparing the reconstructions to the ground truth. We can see how the reconstructions evolve as the photon number increases. At a larger photon number, the 1D cross-section shows a higher contrast and is more similar to the ground truth than at low photon numbers. In general, the sorter-based deconvolution algorithm provides a visibly higher resolution as compared to the conventional deconvolution algorithm, in particular at a low total photon number.

\begin{figure*}[t]
\centering
\includegraphics[width=\linewidth]{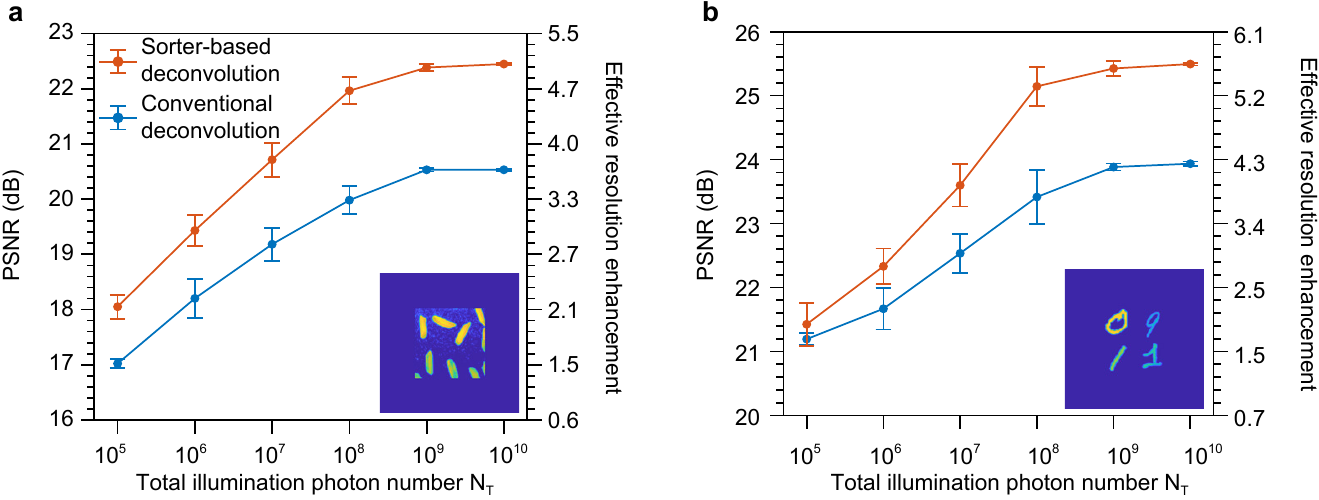}
\caption{The PSNR and effective resolution enhancement as functions of the total photon number $N_T$ for (a) pattern A and (b) pattern B. Both the conventional deconvolution algorithm and the sorter-based deconvolution algorithm are tested for 6 reconstructions with randomly generated Poisson noise. The error bars represent the standard deviation of the PSNR of these trials. The inset shows the corresponding ground truth image.}
\label{fig:figure4}
\end{figure*}

\subsection{Effective resolution enhancement}

In Fig. \ref{fig:figure4} we compare the performance of the conventional deconvolution algorithm to the generalized deconvolution algorithm in terms of PSNR under different $N_T$ for patterns A and B. For each $N_T$, we run the simulation six times with randomly generated Poisson noise to obtain the mean and the standard deviation of the PSNR of the reconstructed images. It can be seen that the generalized deconvolution algorithm based on the mode sorter consistently provides higher PSNR than the conventional confocal approach. Also, the PSNR of the reconstructed image generally increases when $N_T$ increases. Although PSNR is a widely used metric for quantifying the image quality, the PSNR of reconstructed images for different ground truths cannot be compared directly. In addition, the PSNR does not provide an intuitive understanding of the reconstructed resolution. We next translate PSNR to the effective resolution enhancement in order to answer the frequently asked question ``what is the resolution enhancement of your super-resolution method?''. For a ground truth image $W_0$, we blur it with PSFs of different resolutions as $W_{\rm{blur}}=W_0 * M(\delta x)$, where $M(\delta x)$ is the PSF of the conventional confocal microscopy given a particular Rayleigh resolution $\delta x$. We then numerically calculate the PSNR of $W_{\rm{blur}}$ using Eq.~(\ref{eq:PSNR}) to obtain the relation between resolution $\delta x$ and PSNR, i.e. ${\rm{PSNR}}=f(\delta x)$. Therefore, for each reconstructed image $W_r$, we can calculate its effective resolution based on its PSNR via $\delta x_{\rm{eff}}=f^{-1}(\rm{PSNR})$, where $f^{-1}$ is the inverse function of $f$. Hence, the effective resolution enhancement (ERE) can be calculated as
\begin{equation}
\begin{aligned}
{\rm{ERE}} = \delta x_0 / \delta x_{\rm{eff}} ,
\end{aligned}
\end{equation}
and this quantity is shown on the right-hand side axis in Fig.~\ref{fig:figure4}. It can be seen that at the maximum total photon number $N_T= 10^{10}$, the effective resolution enhancement of sorter-based approach is higher than 5.0 for both pattern A and B. We note that the objects used in our simulation have a relatively small space-bandwidth product \cite{bertero2003super}{} because of the limited computational power, which allows for relatively high resolution enhancement. Moreover, the effective resolution enhancement of the sorter-based approach is on average 38\% and 30\% higher than that of the conventional approach for pattern A and pattern B, respectively. We also test nine additional images which are shown in Supplementary Section~4. It can be seen that the mode sorter can provide on average 24\% higher resolution enhancement over the conventional approach for the nine additional objects. We believe that our method can be readily applied to confocal fluorescence microscopy by using a Zernike mode sorter, and the Zernike mode sorter can in principle be experimentally realized by the multi-plane light conversion \cite{labroille2014efficient}. Another potential application of our method is the astronomical imaging where the collected light field is spatially incoherent. However, since the confocal scheme cannot be used in astronomical imaging, the formulas developed here need to be adjusted accordingly to account for the non-confocal scheme used in astronomical imaging.

\section{Conclusion}\label{sec:conclusion}

In conclusion, we generalize the standard RL deconvolution algorithm and apply it to enhance the resolution of sorter-based confocal microscopy. We test our algorithm with general scenes, which has not previously been realized by spatial mode sorting, to the best of our knowledge. The effective resolution enhancement of the sorter-based approach can be as large as $\approx$5.6 when the total photon number in the illumination beam is $N_T= 10^{10}$. For both patterns we test, the average effective resolution enhancement of the sorter-based approach is more than 30\% higher that that of the conventional deconvolution algorithm. Hence, our generalized deconvolution algorithm can achieve robust super-resolution for general scenes compared to the conventional RL deconvolution algorithm. In particular, our generalized deconvolution algorithm allows for super-resolving strongly blurred images of digits, which could be used a front-end to a machine learning-based digit identification task \cite{lecun1998gradient}{}. Furthermore, our method does not require non-classical quantum light sources, and thus our generalized deconvolution algorithm can be potentially useful to applications such as fluorescence microscopy and astronomical imaging. Given the simplicity and generality of our generalized deconvolution algorithm, it is possible to integrate our method with existing quantum or classical super-resolution methods to increase the resolution even further.

\section*{Funding}
National Science Foundation (OMA-1936321, 193632); Defense Advanced Research Projects Agency (D19AP00042); Canada Excellence Research Chairs, Government of Canada; Natural Sciences and Engineering Research Council of Canada; Banting Postdoctoral Fellowship; Universities Space Research Associates (SUBK-20-0002/08103.02); Office of Naval Research (N00014-17-1-2443).

\section*{Acknowledgments}
Y.Z. acknowledges Mankei Tsang for helpful discussions.

\section*{Disclosures}
The authors declare no conflicts of interest.
\\
\\
See Supplement 1 for supporting content.

\clearpage
\begin{center}
\Large{\textbf{Supplemental Document}}
\end{center}

\setcounter{equation}{0} \setcounter{subsection}{0} \setcounter{section}{0}
\setcounter{figure}{0}
\renewcommand{\thesection}{S\arabic{section}}
\renewcommand{\theequation}{S\arabic{equation}}
\renewcommand{\thefigure}{S\arabic{figure}}
\renewcommand{\thetable}{S\arabic{table}}
\renewcommand{\refname}{Supplemental References}
\renewcommand{\bibnumfmt}[1]{[S#1]} 
\renewcommand{\citenumfont}[1]{S#1}

\section{Derivation of Effective PSF of the Conventional Microscopy}
\label{section:DerivationM}

In a conventional confocal microscope, the Airy-disk-shaped PSF of the illumination beam $G(x_1,y_1)$ can be written as
\begin{equation}
\begin{aligned}
 G(x_1,y_1)=N_0 \left| J_1(k\text{NA}r_1)/(\sqrt{\pi}r_1) \right|^2
\end{aligned}
\end{equation}
where $(x_1,y_1)$ are the Cartesian coordinates and $(r_1,\theta_1)$ are the corresponding polar coordinates in the object plane, NA is the numerical aperture, $\lambda$ is wavelength, and $k=2\pi / \lambda$ is the wave number. Here the total energy of the illumination beam is $\iint G(x_1,x_2)dx_1dx_2=N_0$, and thus the illumination beam contains $N_0$ photons. It can be noticed that (see Eq.~(2) in the manuscript):
\begin{equation}
\begin{aligned}
 G(x_1,y_1)=N_0 \frac{k^2 \text{NA}^2}{4\pi} B_{m=0,n=0}(r_1,\theta_1)
\end{aligned}
\end{equation}

To obtain a 2D image, one needs to raster scan the object. This 2D translation is denoted as ($\epsilon, \eta$). The intensity profile of the object is denoted by $W_0(x_1,y_1)$. When the object is excited by the illumination beam, the excited intensity distribution at the object plane is $W_0(x_1-\epsilon,y_1-\eta)G(x_1,y_1)$. For the reflected beam, we assume the imaging system has a magnification of unity, and the intensity PSF at the image plane can be described as
\begin{equation}
\begin{aligned}
\phi(x_2,y_2) = \left| J_1( k \text{NA} r_2) / (\sqrt{\pi}r_2) \right|^2.
\end{aligned}
\end{equation}
where $(x_2,y_2)$ are the Cartesian coordinates at the image plane, $(r_2,\theta_2)$ are the polar coordinates at the image plane, and the energy of $\phi(x_2,y_2)$ is normalized to unity as $\iint  \phi(x_2,y_2) dx_2dy_2=1$. Hence, the resultant image at the image plane can be written as
\begin{equation}
\begin{aligned}
F(x_2,y_2;\epsilon,\eta) = \iint dx_1dy_1 \phi(x_2-x_1,y_2-y_1) W_0(x_1-\epsilon,y_1-\eta) G(x_1,y_1).
\end{aligned}
\end{equation}

In a conventional confocal microscope, we use a pinhole with diameter $D$ at the image plane in front of a bucket detector to measure the photon number. For the conventional confocal microscopy simulation, we use a pinhole of diameter $D=\delta x_0$, which is commonly used in experiments \cite{centonze2006tutorial}{}. Therefore, by raster scanning the object, a 2D image $I_{\rm{con}}(\epsilon,\eta)$ can be obtained by the bucket detector, which can be written as
\begin{equation}
\begin{aligned}
I_{\rm{con}}(\epsilon,\eta) &= \iint_{r_2 \leqslant D/2} dx_2dy_2   F(x_2,y_2;\epsilon,\eta) \\
&=\iint_{r_2 \leqslant D/2} dx_2dy_2  \iint dx_1dy_1 \phi(x_2-x_1,y_2-y_1) W_0(x_1-\epsilon,y_1-\eta) G(x_1,y_1).
\end{aligned}
\label{eq:Error}
\end{equation}

Here we define
\begin{equation}
\begin{aligned}
S(x_1,y_1) &= \iint_{r_2 \leqslant D/2} dx_2  dy_2 \phi(x_2-x_1,y_2-y_1).
\end{aligned}
\end{equation}

Hence, the expression for $I_{\rm{con}}(\epsilon,\eta)$ can be rewritten in the form of a convolution as
\begin{equation}
\begin{aligned}
I(\epsilon,\eta) &=  \iint dx_1dy_1  S(x_1,y_1) G(x_1,y_1) W_0(x_1-\epsilon,y_1-\eta) \\
&= \iint dx_1dy_1  M(x_1,y_1)  W_0(x_1-\epsilon,y_1-\eta) \\
& = M*W_0,
\end{aligned}
\end{equation}
where
\begin{equation}
\begin{aligned}
 M(x_1,y_1)=  S(x_1,y_1) G(x_1,y_1) .
\end{aligned}
\end{equation}

\section{Derivations of Effective PSF of the Mode Sorter-Based Microscopy }
\label{DerivationQHB}

For our sorter-based method, we use a Zernike mode sorter to perform spatial mode decomposition. The expression for the normalized Zernike modes in the Fourier plane is
\begin{equation}
Z^m_{n}(r_p,\theta_p) = \sqrt{\frac{2(n+1)}{\epsilon_m \pi}} R^{|m|}_n (r_p) \sin(m \theta_p + \frac{\pi}{2}\cdot {\mathcal{H}}(m)),
\end{equation}
where $(r_p,\theta_p)$ are the scaled, dimensionless polar coordinates at the Fourier plane with $0 \leqslant r_p \leqslant 1$; $\epsilon_m=2$ if $m=0$ and $\epsilon_m=1$ if $m\neq 0$; ${\mathcal{H}}(m)$ is the Heaviside function where ${\mathcal{H}}(m)=1$ if $m \geqslant 0$ and ${\mathcal{H}}(m)=0$ if $m < 0$; $R^{|m|}_n$ is the radial polynomial \cite{janssen2011new}{}. Consider a point source located at $(x_1,y_1)$. Then the electric field on the scaled pupil plane is $\frac{1}{\sqrt{\pi}}\exp ( i k \text{NA} (x_1 x_p+y_1 y_p) )$, where $(x_p,y_p)$ is the dimensionless Cartesian coordinate at the Fourier plane with $x_p^2+y_p^2 \leqslant 1$. In the polar coordinate, this field can be rewritten as $\frac{1}{\sqrt{\pi}}\exp ( i k \text{NA} r_1 r_p \cos (\theta_1-\theta_p) )$. Now we project this field to the Zernike modes $Z^m_n(r_p,\theta_p)$, and the overlap integral is \cite{janssen2011new}{}
\begin{equation}
\begin{aligned}
B_{mn}(x_1,y_1)  &=\left\lvert \int_0^1 r_pdr_p \int_0^{2\pi}d\theta_p   \frac{1}{\sqrt{\pi}}\exp ( i k \text{NA} r_1 r_p \cos (\theta_1-\theta_p) ) \cdot  Z^m_{n}(r_p,\theta_p)   \right\lvert^2\\
&=\frac{8(n+1)}{\epsilon_m } \frac{J^2_{n+1}(k \text{NA} r_1)}{(k\text{NA}r_1)^2}\sin^2(m\theta_1+\frac{\pi}{2}\cdot {\rm{H}}(m)).
\end{aligned}
\end{equation}
It can be seen that the overlap integral coincides with the Fourier transform of Zernike modes. Therefore, when the object $W_0(x_1-\epsilon,y_1-\eta)$ is excited by the illumination beam $G(x_1,y_1)$, the sorter output is
\begin{equation}
\begin{aligned}
H_{mn}(\epsilon,\eta)  &= \iint dx_1 dy_1  B_{mn}(x_1,y_1) G(x_1,y_1) W_0(x_1-\epsilon,y_1-\eta) \\
& =  \iint dx_1 dy_1  Q_{mn}(x_1,y_1)  W_0(x_1-\epsilon,y_1-\eta) \\
&= Q_{mn}*W_0,
\end{aligned}
\end{equation}
where
\begin{equation}
\begin{aligned}
Q_{mn}(x_1,y_1) &=B_{mn}(x_1,y_1) \cdot G(x_1,y_1) \\
&= N_0 \frac{k^2 \text{NA}^2}{4\pi} B_{m=0,n=0}(x_1,y_1) \cdot B_{mn}(x_1,y_1) .
\end{aligned}
\end{equation}

\clearpage

\section{Quality of Iterative Reconstructions}
\label{Iterations}

\begin{figure}[h]
\centering
\includegraphics[width=\linewidth]{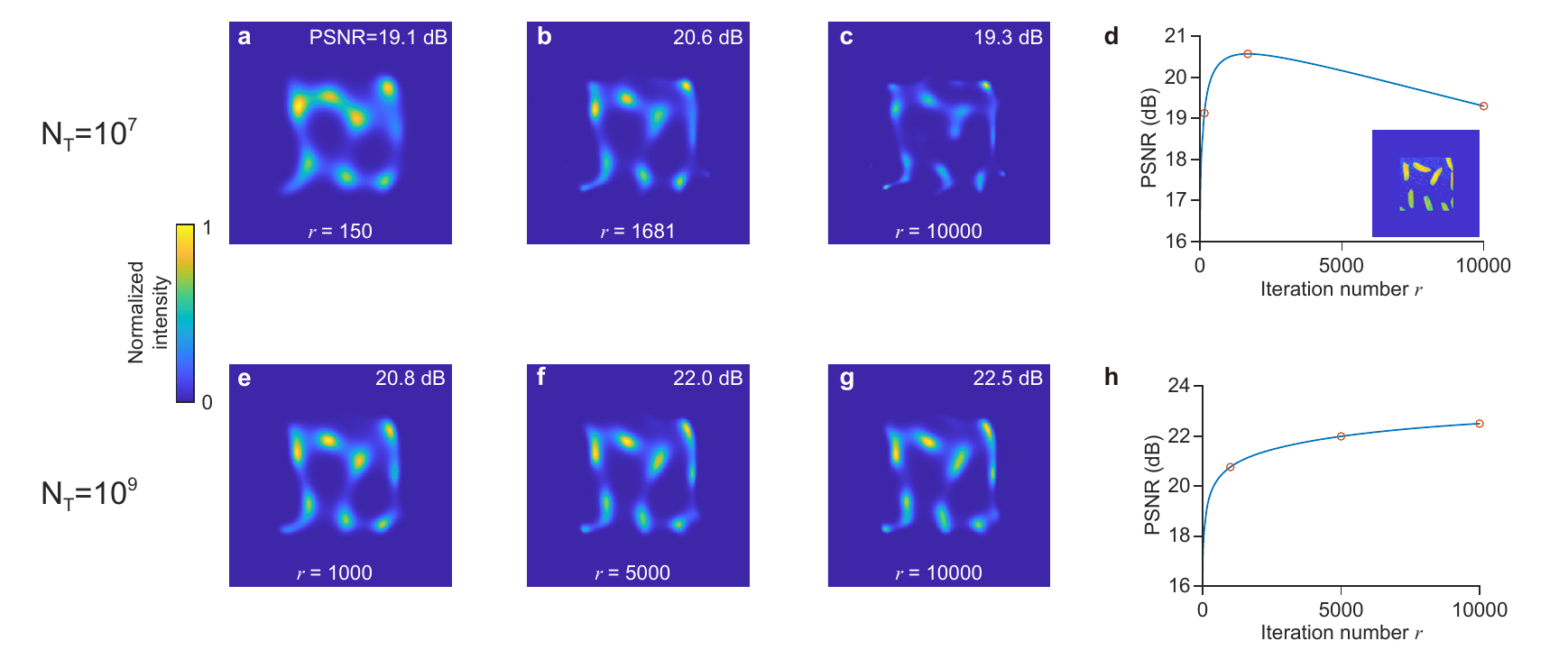}
\caption{(a-c) The reconstructed image at different iteration numbers by the sorter-based method when the total photon number $N_T= 10^7$. The corresponding PSNR as a function of iteration number $r$ is shown in (d). The noise in the image can be amplified when the iteration number exceeds the optimal value. The inset shows the ground truth image. (e-g) The reconstructed image at different iteration numbers by the sorter-based method when the total photon number $N_T=10^9$. The corresponding PSNR as a function of iteration number $r$ is shown in (h).}
\label{fig:MoreData2}
\end{figure}

We allow the algorithm to run until it reaches a maximum iteration number $N_{\rm{ite}}= 10^4$, which is limited by time and computational power constraints. We monitor the PSNR at each iteration $r$ and choose the maximum PSNR for $1 \leqslant r \leqslant N_{\rm{ite}}$ for each implementation. In general, the PSNR increases with increasing iteration number $r$, as shown in Fig.~\ref{fig:MoreData2}(h).  However, noisy data can cause the PSNR to decrease with increasing $r$, as the noise can be amplified when the iteration number is large. This effect is shown in Fig.~\ref{fig:MoreData2}(d). In Fig.~\ref{fig:MoreData2}(a-c) we see the progression of the image quality as the iteration number is increased. Fig.~\ref{fig:MoreData2}(e-g) illustrates the same progression when the total photon number $N_T$ in the illumination beam increases. These images are reconstructions of pattern A.

\begin{figure}[t]
\centering
\includegraphics[width=\linewidth]{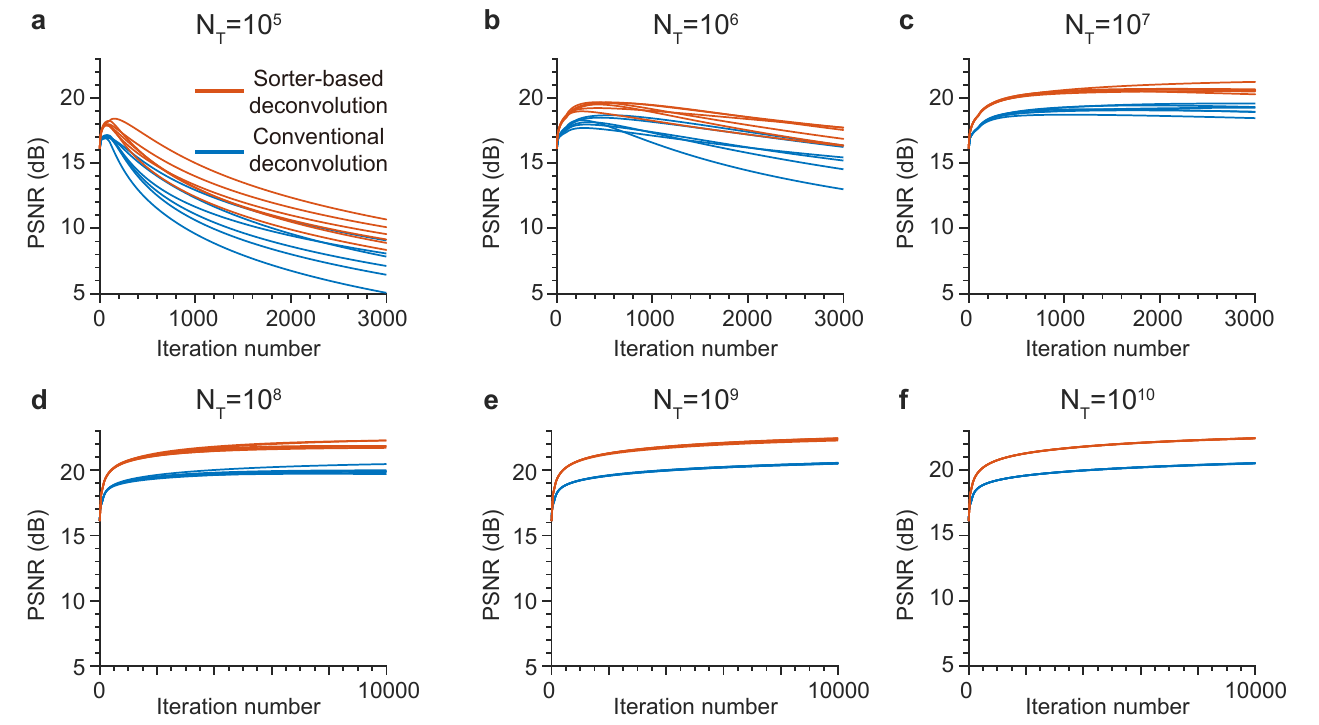}
\caption{The relation between the PSNR and the iteration number with different $N_T$ for the conventional and sorter-based deconvolution algorithm. The value of $N_T$ is shown on the top of each graph.}
\label{fig:PSNR_Compare}
\end{figure}

In practical applications where the ground truth is unavailable, a stopping criterion is needed to stop the iterative algorithm. The simplest (and perhaps the most widely used) stopping criterion is to manually specify a maximum iteration number. To illustrate the effect of the manually specified stopping criterion, we show the relation between the PSNR and the iteration number for the conventional deconvolution algorithm and the sorter-based deconvolution algorithm with different $N_T$ in Fig.~\ref{fig:PSNR_Compare}. For each $N_T$, we generate six shot-noise-limited images with randomly generated Poisson noise. Therefore, we have 12 curves in each graph in total. It can be seen that the sorter-based deconvolution algorithm outperforms the conventional deconvolution algorithm for an arbitrary iteration number. To achieve the optimal performance, the user should choose a smaller iteration number when $N_T$ is small. The results also show that both the conventional and sorter-based deconvolution algorithms have a similar dependence on the iteration number, and thus the PSNR improvement is almost independent of the chosen stopping criterion.

\clearpage

\section{Performance characterization for Additional Images}
\label{AddImages}

\begin{figure}[h]
\centering
\includegraphics[width=\linewidth]{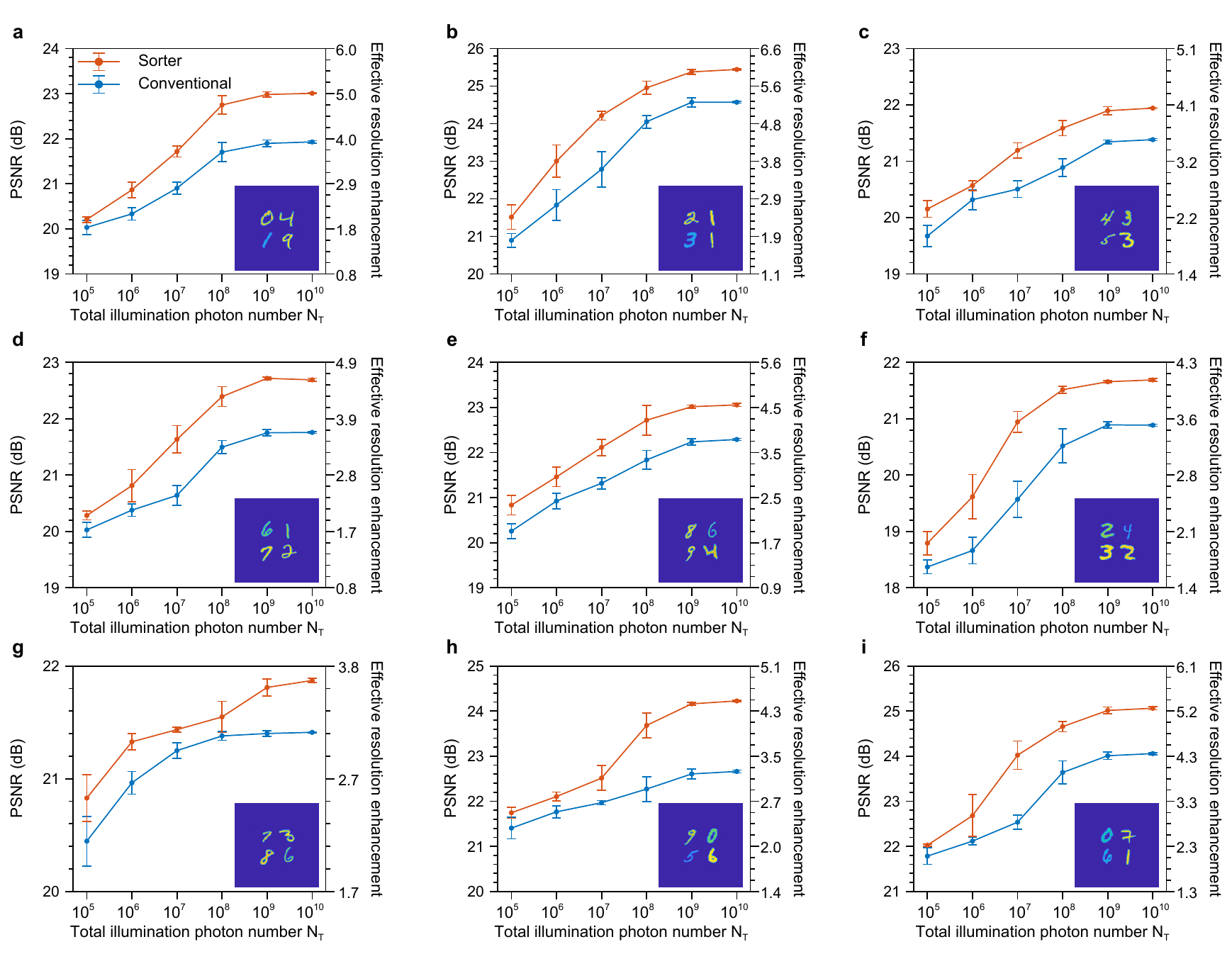}
\caption{(a-i) Performance characterization for nine different objects. The object image is shown as an inset in each figure.}
\label{fig:MoreData}
\end{figure}

The main goal of our approach is to super-resolve an object of arbitrary geometry using a mode sorter, as to the best of our knowledge this has not yet been accomplished. In order to test our algorithm, we fed it many images to reconstruct. In Fig.~\ref{fig:MoreData} we compare the performance of the conventional algorithm to the generalized algorithm in terms of PSNR and the effective resolution enhancement (ERE). Among all the data points shown in the inset of Fig.~\ref{fig:MoreData} (a-i),
the PSNR enhancement PSNR$_{\text{sorter}}-$PSNR$_{\text{conventional}}$ has a minimum value of 0.17 dB, a maximum value of 1.73 dB, and an average value of 0.82 dB. The effective resolution enhancement ratio ERE$_{\text{sorter}}/$ERE$_{\text{conventional}}$ has a minimum value of 107\%, a maximum value of 150.5\%, and an average value of 124.2\%.

\begin{thebibliography}{10}
\newcommand{\enquote}[1]{``#1''}

\bibitem{rayleigh1879xxxi}
L.~Rayleigh, \enquote{{XXXI}. {Investigations} in optics, with special
  reference to the spectroscope,} {\protect\JournalTitle{Philos. Mag. Ser.}}
  \textbf{8}, 261--274 (1879).

\bibitem{hell1994breaking}
S.~W. Hell and J.~Wichmann, \enquote{Breaking the diffraction resolution limit
  by stimulated emission: stimulated-emission-depletion fluorescence
  microscopy,} {\protect\JournalTitle{Opt. Lett.}} \textbf{19}, 780--782
  (1994).

\bibitem{betzig2006imaging}
E.~Betzig, G.~H. Patterson, R.~Sougrat, O.~W. Lindwasser, S.~Olenych, J.~S.
  Bonifacino, M.~W. Davidson, J.~Lippincott-Schwartz, and H.~F. Hess,
  \enquote{Imaging intracellular fluorescent proteins at nanometer resolution,}
  {\protect\JournalTitle{Science}} \textbf{313}, 1642--1645 (2006).

\bibitem{rust2006sub}
M.~J. Rust, M.~Bates, and X.~Zhuang, \enquote{Sub-diffraction-limit imaging by
  stochastic optical reconstruction microscopy ({STORM}),}
  {\protect\JournalTitle{Nat. Methods}} \textbf{3}, 793--796 (2006).

\bibitem{tsang2009quantum}
M.~Tsang, \enquote{Quantum imaging beyond the diffraction limit by optical
  centroid measurements,} {\protect\JournalTitle{Phys. Rev. Lett.}}
  \textbf{102}, 253601 (2009).

\bibitem{shin2011quantum}
H.~Shin, K.~W.~C. Chan, H.~J. Chang, and R.~W. Boyd, \enquote{Quantum spatial
  superresolution by optical centroid measurements,}
  {\protect\JournalTitle{Phys. Rev. Lett.}} \textbf{107}, 083603 (2011).

\bibitem{rozema2014scalable}
L.~A. Rozema, J.~D. Bateman, D.~H. Mahler, R.~Okamoto, A.~Feizpour, A.~Hayat,
  and A.~M. Steinberg, \enquote{Scalable spatial superresolution using
  entangled photons,} {\protect\JournalTitle{Phys. Rev. Lett.}} \textbf{112},
  223602 (2014).

\bibitem{unternahrer2018super}
M.~Untern{\"a}hrer, B.~Bessire, L.~Gasparini, M.~Perenzoni, and A.~Stefanov,
  \enquote{Super-resolution quantum imaging at the {Heisenberg} limit,}
  {\protect\JournalTitle{Optica}} \textbf{5}, 1150--1154 (2018).

\bibitem{toninelli2019resolution}
E.~Toninelli, P.-A. Moreau, T.~Gregory, A.~Mihalyi, M.~Edgar, N.~Radwell, and
  M.~Padgett, \enquote{Resolution-enhanced quantum imaging by centroid
  estimation of biphotons,} {\protect\JournalTitle{Optica}} \textbf{6},
  347--353 (2019).

\bibitem{tenne2019super}
R.~Tenne, U.~Rossman, B.~Rephael, Y.~Israel, A.~Krupinski-Ptaszek,
  R.~Lapkiewicz, Y.~Silberberg, and D.~Oron, \enquote{Super-resolution
  enhancement by quantum image scanning microscopy,}
  {\protect\JournalTitle{Nat. Photon.}} \textbf{13}, 116--122 (2019).

\bibitem{schwartz2013superresolution}
O.~Schwartz, J.~M. Levitt, R.~Tenne, S.~Itzhakov, Z.~Deutsch, and D.~Oron,
  \enquote{Superresolution microscopy with quantum emitters,}
  {\protect\JournalTitle{Nano Lett.}} \textbf{13}, 5832--5836 (2013).

\bibitem{tsang2016quantum}
M.~Tsang, R.~Nair, and X.-M. Lu, \enquote{Quantum theory of superresolution for
  two incoherent optical point sources,} {\protect\JournalTitle{Phys. Rev. X}}
  \textbf{6}, 031033 (2016).

\bibitem{yang2016far}
F.~Yang, A.~Tashchilina, E.~S. Moiseev, C.~Simon, and A.~I. Lvovsky,
  \enquote{Far-field linear optical superresolution via heterodyne detection in
  a higher-order local oscillator mode,} {\protect\JournalTitle{Optica}}
  \textbf{3}, 1148--1152 (2016).

\bibitem{paur2016achieving}
M.~Pa{\'u}r, B.~Stoklasa, Z.~Hradil, L.~L. S{\'a}nchez-Soto, and J.~Rehacek,
  \enquote{Achieving the ultimate optical resolution,}
  {\protect\JournalTitle{Optica}} \textbf{3}, 1144--1147 (2016).

\bibitem{tham2017beating}
W.-K. Tham, H.~Ferretti, and A.~M. Steinberg, \enquote{Beating {Rayleigh's}
  curse by imaging using phase information,} {\protect\JournalTitle{Phys. Rev.
  Lett.}} \textbf{118}, 070801 (2017).

\bibitem{tang2016fault}
Z.~S. Tang, K.~Durak, and A.~Ling, \enquote{Fault-tolerant and finite-error
  localization for point emitters within the diffraction limit,}
  {\protect\JournalTitle{Opt. Express}} \textbf{24}, 22004--22012 (2016).

\bibitem{zhou2019quantum}
Y.~Zhou, J.~Yang, J.~D. Hassett, S.~M.~H. Rafsanjani, M.~Mirhosseini, A.~N.
  Vamivakas, A.~N. Jordan, Z.~Shi, and R.~W. Boyd, \enquote{Quantum-limited
  estimation of the axial separation of two incoherent point sources,}
  {\protect\JournalTitle{Optica}} \textbf{6}, 534--541 (2019).

\bibitem{tsang2018subdiffraction}
M.~Tsang, \enquote{Subdiffraction incoherent optical imaging via spatial-mode
  demultiplexing: Semiclassical treatment,} {\protect\JournalTitle{Phys. Rev.
  A}} \textbf{97}, 023830 (2018).

\bibitem{albarelli2019evaluating}
F.~Albarelli, J.~F. Friel, and A.~Datta, \enquote{Evaluating the {Holevo}
  {Cram{\'e}r-Rao} bound for multiparameter quantum metrology,}
  {\protect\JournalTitle{Phys. Rev. Lett.}} \textbf{123}, 200503 (2019).

\bibitem{tsang2019semiparametric}
M.~Tsang, \enquote{Semiparametric estimation for incoherent optical imaging,}
  {\protect\JournalTitle{Phys. Rev. Research}} \textbf{1}, 033006 (2019).

\bibitem{zhou2019modern}
S.~Zhou and L.~Jiang, \enquote{Modern description of {Rayleigh's} criterion,}
  {\protect\JournalTitle{Phys. Rev. A}} \textbf{99}, 013808 (2019).

\bibitem{yang2019optimal}
J.~Yang, S.~Pang, Y.~Zhou, and A.~N. Jordan, \enquote{Optimal measurements for
  quantum multiparameter estimation with general states,}
  {\protect\JournalTitle{Phys. Rev. A}} \textbf{100}, 032104 (2019).

\bibitem{tsang2017subdiffraction}
M.~Tsang, \enquote{Subdiffraction incoherent optical imaging via spatial-mode
  demultiplexing,} {\protect\JournalTitle{New J. Phys.}} \textbf{19}, 023054
  (2017).

\bibitem{vrehavcek2017multiparameter}
J.~{\v{R}}eha{\v{c}}ek, Z.~Hradil, B.~Stoklasa, M.~Pa{\'u}r, J.~Grover,
  A.~Krzic, and L.~S{\'a}nchez-Soto, \enquote{Multiparameter quantum metrology
  of incoherent point sources: towards realistic superresolution,}
  {\protect\JournalTitle{Phys. Rev. A}} \textbf{96}, 062107 (2017).

\bibitem{lu2018quantum}
X.-M. Lu, H.~Krovi, R.~Nair, S.~Guha, and J.~H. Shapiro,
  \enquote{Quantum-optimal detection of one-versus-two incoherent optical
  sources with arbitrary separation,} {\protect\JournalTitle{npj Quantum Inf.}}
  \textbf{4}, 1--8 (2018).

\bibitem{prasad2020quantum}
S.~Prasad, \enquote{Quantum limited source localization and pair
  superresolution in two dimensions under finite-emission bandwidth,}
  {\protect\JournalTitle{Phys. Rev. A}} \textbf{102}, 033726 (2020).

\bibitem{bonsma2019realistic}
K.~A. Bonsma-Fisher, W.-K. Tham, H.~Ferretti, and A.~M. Steinberg,
  \enquote{Realistic {sub-Rayleigh} imaging with phase-sensitive measurements,}
  {\protect\JournalTitle{New J. Phys.}} \textbf{21}, 093010 (2019).

\bibitem{tsang2020efficient}
M.~Tsang, \enquote{Efficient superoscillation measurement for incoherent
  optical imaging,} {\protect\JournalTitle{arXiv:2010.11084}}  (2020).

\bibitem{peng2020generalization}
L.~Peng and X.-M. Lu, \enquote{Generalization of {Rayleigh's} curse on
  parameter estimation with incoherent sources,}
  {\protect\JournalTitle{arXiv:2011.07897}}  (2020).

\bibitem{richardson1972bayesian}
W.~H. Richardson, \enquote{Bayesian-based iterative method of image
  restoration,} {\protect\JournalTitle{J. Opt. Soc. Am.}} \textbf{62}, 55--59
  (1972).

\bibitem{lucy1974iterative}
L.~B. Lucy, \enquote{An iterative technique for the rectification of observed
  distributions,} {\protect\JournalTitle{Astron. J.}} \textbf{79}, 745 (1974).

\bibitem{airy1835diffraction}
G.~B. Airy, \enquote{On the diffraction of an object-glass with circular
  aperture,} {\protect\JournalTitle{Trans. of the Cambridge Philosoph. Soc.}}
  \textbf{5}, 283--291 (1835).

\bibitem{yu2018quantum}
Z.~Yu and S.~Prasad, \enquote{Quantum limited superresolution of an incoherent
  source pair in three dimensions,} {\protect\JournalTitle{Phys. Rev. Lett.}}
  \textbf{121}, 180504 (2018).

\bibitem{nasrollahi2014super}
K.~Nasrollahi and T.~B. Moeslund, \enquote{Super-resolution: a comprehensive
  survey,} {\protect\JournalTitle{Mach. Vision Appl.}} \textbf{25}, 1423--1468
  (2014).

\bibitem{herbert1990statistical}
T.~Herbert, \enquote{Statistical stopping criteria for iterative maximum
  likelihood reconstruction of emission images,} {\protect\JournalTitle{Phys.
  Med. Biol.}} \textbf{35}, 1221 (1990).

\bibitem{lecun1998gradient}
Y.~LeCun, L.~Bottou, Y.~Bengio, and P.~Haffner, \enquote{Gradient-based
  learning applied to document recognition,} {\protect\JournalTitle{Proc. of
  the IEEE}} \textbf{86}, 2278--2324 (1998).

\bibitem{bertero2003super}
M.~Bertero and P.~Boccacci, \enquote{Super-resolution in computational
  imaging,} {\protect\JournalTitle{Micron}} \textbf{34}, 265--273 (2003).

\bibitem{labroille2014efficient}
G.~Labroille, B.~Denolle, P.~Jian, P.~Genevaux, N.~Treps, and J.-F. Morizur,
  \enquote{Efficient and mode selective spatial mode multiplexer based on
  multi-plane light conversion,} {\protect\JournalTitle{Opt. Express}}
  \textbf{22}, 15599--15607 (2014).

\end{thebibliography}

\begin{thebibliography}{1}
\newcommand{\enquote}[1]{``#1''}

\bibitem{centonze2006tutorial}
V.~Centonze and J.~B. Pawley, \enquote{Tutorial on practical confocal
  microscopy and use of the confocal test specimen,} in \emph{Handbook of
  biological confocal microscopy,}  (Springer, 2006), pp. 627--649.

\bibitem{janssen2011new}
A.~Janssen, \enquote{New analytic results for the {Zernike} circle polynomials
  from a basic result in the {Nijboer-Zernike} diffraction theory,}
  {\protect\JournalTitle{J. Eur. Opt. Soc.}} \textbf{6} (2011).

\end{thebibliography}
\end{document}